\documentstyle [12pt] {article}
\newcommand{\cs}[3]{{{#3} \brace {#1 #2}}}
\begin{document}
\title {Spinning Equations for Objects of Some Classes in Finslerian Geometry}
\maketitle

\begin{center}
{\bf {Magd E. Kahil{\footnote{ Faculty of Engineering, Modern Sciences and Arts University, Giza, Egypt  \\
e.mail: mkahil@msa.eun.eg}}} }
\footnote{Egyptian Relativity Group. Cairo, Egypt}
\end{center}

\begin{abstract}
     Spinning equations of Finslerian gravity, the counterpart of Mathisson-Papapetrou Spinning equations of motion are obtained. Two approaches of Finslerian geometries  are illustrated, and their corresponding spinning equations. The significance of nonlinear connection and its relevance on spinning equations and their deviations ones are  examined.
\end{abstract}
\section{Introduction}
The problem of motion of test particle has been discussed by many authors in Finslerian geometry as an introductory to express the behavior of particles defined within the frame work of this geometry [1]. This type of achievement is always compared with its counterpart in The Riemaniann geometry. It is worth mentioning that the most important significance of Finslerian geometry is describing at each point in its manifold by coordinates and its direction, leads to replace manifolds with tangent bundles [2]. Accordingly, a new appeared quantities may be expected, able to interpret it from physical point of view. This type of technicality is due to implementing  the concept of geometrization of physics.

In our study, we are going to obtain the spinning equations due to Rund's approach of  the orthodox Finsler space [3] and the Cartan approach which is based on introducing the role of nonlinear connection [4], having the role of gauge potential [5] ,which is absent in the context of the usual Riemannian geometry.

The aim of this work is searching for a  suitable Lagrangian function, inspired from the famous  Bazanski Lagrangian [6] as expressed in previous works [7-9].

Thus, it is vital to derive their corresponding geodesic and geodesic deviation to obtain  the  version  of the spinning object for short, using a specific type of transformation explained previously in our present work [9]. Consequently, this step will enable us to make sure of the reliability of the obtained equation. This will guide us to  determine its corresponding Lagrangian function, a step toward generalization,  to describe the spinning equation of  different  objects .  Also, we  present   a technique of commutation to obtain their deviation equation rather than the traditional method of Bazanski [6], which is based on  the variation with respect to four velocity  velocity, as well as using some identities  to preserve the appearance of covariance in the system of equations  [7-9] . The advantage of this method may give rise to examine the case of rate of change of spinning object and its associated deviation equation.

\section{Finlerian Geometry: A Brief Introduction}
Finslerian spaces are n-dimensional manifolds, the distance between two neighboring points $P(x^{i})$ and $Q(x^{i}+ dx^{i})$ is given by the line element [1]
$$
ds = F (x^{i}, dx^{i}), i= 1,2,3,...n,
$$
where $F$ is a function in both coordinates $x^{i}$ and directions $\dot{x}^{i} = \frac{d x^{i}}{d s}$ .\\

The metric of the Finsler space is defined as
\begin{equation}
g_{\mu \nu}  = \frac{\partial^{2} F^{2}(x, \dot{x})}{\partial \dot{x}^{\mu}\partial \dot{x}^{\nu} }
\end{equation}
such that
\begin{equation}
g_{\mu \nu}  = g_{\mu \nu}dx^{\mu}\bigotimes dx^{\nu}
\end{equation}

A third order tensor , the Cartan tensor is defined as follows
\begin{equation}
C_{(x, \dot{x})} = C_{\mu \nu \rho}dx^{\mu}\bigotimes dx^{\nu}\bigotimes dx^{\rho}
\end{equation}
such that
\begin{equation}
 C_{\mu \nu \rho} = \frac{1}{2} \frac{\partial g_{\mu \nu}}{\partial \dot{x}^{\nu} \partial \dot{x}^{\rho}} ,
  \end{equation}
i.e.
\begin{equation}
 C_{\mu \nu \rho} = \frac{1}{2} \frac{\partial^{3} F^{2} }{\partial\dot{x}^{\mu} \partial \dot{x}^{\nu} \partial \dot{x}^{\rho}} .
  \end{equation}

The Cartan tensor measures the deviation from the Riemannian manifold, if  $C_{\mu \nu \rho}=0$ the space becomes a Riemannian one [2]
and the metric becomes independent of $\dot x^{\rho}$ , the metric $g_{\mu \nu}(x)$ becomes function of the coordinates only. It is well known that the geodesic equation of the Finsler geometry is described by
\begin{equation}
\frac{d^{2} x^{\mu}}{d S^2 } + 2 G^{\mu} =0,
\end{equation}
where the spray $G^{\mu}$ is defined as follows
$$
G^{\mu} = \frac{1}{4}g^{\mu \nu} [\dot{x}^{\sigma} \frac{\partial^{2}F^{2}}{\partial x^{\nu}\partial \dot{x}^{\sigma}} - \frac{\partial{F^{2}}}{\partial x^{\nu}}].
$$
 From this point of view, it is well known in the Fnslerian geometry, there are two different types of connections linear connections as expressed in the Riemannian geometry with slight modifications, The Rund and Finsler-Cartan approaches. \\

 \subsection{Finsler geometry: The Rund Approach}
  Rund defined $\bar{\delta}$-derivative [3] in which
 \begin{equation}
 \frac{\bar{\delta} A^{\mu} }{\bar{\delta} S} = \frac{d A^{\mu}}{dt} + {}^*\Gamma^{\mu}_{\nu \rho}A^{\nu}\dot{x}^{\rho}.
 \end{equation}
The partial $\delta$-derivative is defined as follows,
 \begin{equation}
A^{\mu}_{\bar{;} \nu} = \frac{\partial A^{\mu}}{\partial x^{\nu}} + {}^{*}\Gamma^{\mu}_{\nu \rho} A^{\nu} \dot{x}^{\rho},
 \end{equation}
  such that
  \begin{equation}
 \frac{\bar{\delta} A^{\mu} }{\delta s} = A^{\mu}_{\bar{;} \nu} \frac{d x^{\nu}}{d s}
\end{equation}
such that
 $$
 {}^*\Gamma^{\mu}_{\nu \rho} = \cs{\nu}{\rho}{\mu} - g^{\mu \alpha} [ C_{\rho \alpha \beta}\Gamma^{\beta}_{ \nu} \epsilon + C_{\nu \alpha\beta} \Gamma^{\beta}_{\rho \epsilon} - C_{\nu \rho \beta} \Gamma^{\beta}_{\alpha \epsilon}] \dot{x}^{\epsilon}
 $$

 $$
 \Gamma^{\mu}_{\nu \rho} = \gamma^{\mu}_{\nu \rho} \cs{\nu}{\rho}{\mu} - C^{\mu}_{\nu \delta}\cs{\lambda}{\rho}{\delta} \dot{x}^{\lambda},
 $$

 $$
 C^{\mu}_{\nu \delta}= g^{\mu \delta}C_{\nu \delta \rho} = g^{\mu \delta}(\frac{\partial g_{\nu \rho}}{\partial \dot{x}^{\delta}})
 $$
 Equation (7) reduces to Riemannian geometry if $C_{\nu \delta \rho} \equiv 0$ .

  The curvature tensor is defined  to the commutation law as follows,
  \begin{equation}
  A^{\mu}_{\bar{;} \nu \rho}- A^{\mu}_{\bar{;} \rho \nu} = K^{\mu}_{\nu \rho \delta}A^{\delta},
  \end{equation}
 where $ K^{\mu}_{\nu \rho \delta}$ the curvature tensor defined as  follows.
 $$
 K^{\mu}_{\nu \rho \sigma} = ({}^*\Gamma^{\mu}_{\nu \rho, \sigma } - \frac{\partial {}^*\Gamma^{\mu}_{\nu  \rho} }{\partial \dot{x}^{\delta}} \frac{ \partial G^{\delta}}{\partial \dot{x}^{\sigma}} )- ({}^*\Gamma^{\mu}_{\nu \sigma, \rho } - \frac{\partial {}^*\Gamma^{\mu}_{\nu  \sigma} }{\partial \dot{x}^{\delta}} \frac{ \partial G^{\delta}}{\partial \dot{x}^{\rho}})+ {}^*\Gamma^{\lambda}_{\nu \rho }{}^*\Gamma^{\mu}_{\lambda  \sigma } - + {}^*\Gamma^{\lambda}_{\nu \sigma }{}^*\Gamma^{\mu}_{\lambda  \rho },
 $$
 where $$ G^{\mu} = \frac{1}{2} \cs{\nu}{\rho}{\mu} \dot{x}^{\nu}\dot{x}^{\nu}, $$

\subsection{ Geodesic and geodesic deviation in Finsler Space: The Rund Approach}

 Equations of  geodesic and geodesic deviation  associated  with the  Finsler geometry in this approach may be performed by applying the Euler-Lagrange equation using suggested the Lagrangian function.
\begin{equation}
L = g_{\mu \nu} (x, \dot{x}) {U}^{\alpha} \frac{\bar{\delta}{\Psi}^{\beta}}{ \bar{\delta} S},
\end{equation}
 where $\Psi^{mu}$ is S- dependent deviation vector associated with one parameter of curves $x^{\mu}(S, \epsilon)$  such that [11]

 $$
 \Psi^{\mu} = \tau \frac{\partial x^{\mu}}{\partial \tau} |_{\tau =0} .
 $$
Taking the variation with respect to ${\Psi}^{\mu}$ we obtain,
\begin{equation}
 \frac{\bar{\delta} {U}^{\beta}}{ \bar{\delta} S} = 0.
\end{equation}
While, their deviation equations can be obtained in a similar way as obtained in [12] using the following commutation relation
 \begin{equation}
 \frac{\bar{\delta} U^{\mu}}{\bar{\delta} S} = \frac{\delta \Psi}{\partial \tau}
 \end{equation}
Using (4)and (7) after some manipulations one obtains,
\begin{equation}
 \frac{\bar{\delta}^{2} {\Psi}^{\beta}}{ \bar{\delta} S^2} = K^{\alpha}_{\mu \nu \sigma} U^{\mu} U^{\nu} \Psi^{\sigma} .
\end{equation}

\section{Spinning equation in Finslerian Geometry :The Rund Approach}
Using the following type of transformation [13]
 \begin{equation}
 {V}^{\mu} = U^{\mu} + \beta \frac{\bar{\delta}\Psi^{\mu}}{\bar{\delta} S}
 \end{equation}
 where , $\beta$ is a parameter. \\
 Taking the absolute $\delta$ derivative on both sides
\begin{equation}
 \frac{\bar{\delta}{V}^{\mu}}{\bar{\delta}\rho} = \frac{\bar{\delta}}{\bar{\delta} S }(U^{\mu} + \beta \frac{\bar{\delta} \Psi^{\mu}}{\bar{\delta} S}) \frac{d S}{d \rho}
 \end{equation}
Substituting from geodesic and geodesic deviation equations and using the following definition [10]
\begin{equation}
 S^{\mu \nu} = \sigma [(U^{\mu}\Psi^{\nu})- (U^{\nu}\Psi^{\mu}) ]
\end{equation}
in which $\beta = \frac{\sigma}{m}$ , where , $S^{\mu \nu}$ is the spin tensor, $\sigma$ is the magnitude of spin and $m$ is mass of the object to get
\begin{equation}
\frac{\bar{\delta} V^{\alpha}}{{\delta}\rho} = \frac{1}{2m} K^{\alpha}_{\beta \gamma \epsilon} S^{\gamma \epsilon} V^{\beta}
\end{equation}
which is the analogous version to the Papertrou Equation [14] of spinning objects for short.

\subsection{Spinning and Spinning Deviation equation: The Rund Approach}
{\underline{{i} Case $ P^{\mu} = m U^{\mu}$}}
\begin{equation}
L= g_{\mu \nu} (x, \dot{x}) U^{\mu} \frac{\bar{\delta} \Phi^{\nu}}{\bar{\delta}S} + S_{\mu \nu}\frac{\bar{\delta} \Psi^{\mu \nu}}{\bar{\delta} S} + \frac{1}{2m} K_{\mu \nu \rho \sigma} S^{\rho \sigma}U^{\nu} \Psi^{\mu}
 \end{equation}

By taking the variation with respect to $\Psi^{\alpha}$ and $\Psi{\alpha \beta}$ to get

\begin{equation}
 \frac{\bar{\delta}U^{\beta}}{ \bar{\delta} S} = \frac{1}{2m} K^{\alpha}_{\mu \nu \sigma}S^{\nu \sigma} U^{\mu} ,
\end{equation}
and
\begin{equation}
 \frac{\bar{\delta}S^{\alpha \beta}}{ \bar{\delta} S} = 0 ,
\end{equation}

In order to obtain their corresponding deviation equation, we take the covariant delta derivative with respect to the path $\tau$ on both sides and using the following conditions
\begin{equation}
\frac{\bar{\delta}\Psi^{\beta}}{ \bar{\delta} S} = \frac{\bar{\delta}U^{\beta}}{ \bar{\delta} \tau},
\end{equation}
and
\begin{equation}
 \frac{\bar{\delta}}{\bar{\delta} S }\frac{\bar{\delta} {A}}{\bar{\delta} \tau }- \frac{\bar{\delta}}{\bar{\delta} \tau } \frac{\bar{\delta} {A}}{\bar{\delta} S } = K^{\alpha}_{\beta \gamma \epsilon} U^{\beta} U^{\gamma} \Psi^{\epsilon},
\end{equation}

to obtain the set of equations

\begin{equation}
 \frac{\bar{\delta}^2 \Psi^{\beta}}{\bar{\delta} S^2} = K^{\alpha}_{\mu \nu \sigma}U^{\mu} U^{\nu}\Psi^{\sigma}  + \frac{1}{2m} (K^{\alpha}_{\mu \nu \sigma}S^{\nu \sigma} U^{\mu})_{\bar{;} \rho} \Psi^{\rho} ,
\end{equation}
and
\begin{equation}
 \frac{\bar{\delta}^2 \Psi^{\alpha \beta}}{ \bar{\delta} S^2} = S^{[ \alpha \mu} K^{\beta ]}_{\mu \nu \sigma} U^{\nu}\Psi^{\sigma} .  
\end{equation}
{\underline{{ii} Case $ P^{\mu} \neq  m U^{\mu}$}}\\
In this case we are going to derive the  corresponding equations of motion and their deviation ones by proposing the following Lagrangian function
\begin{equation}
L= g_{\mu \nu} (x, \dot{x}) U^{\mu} \frac{\bar{\delta} \Phi^{\nu}}{\bar{\delta}S} + S_{\mu \nu}\frac{\bar{\delta} \Psi^{\mu \nu}}{\bar{\delta} S} + \frac{1}{2m} K_{\mu \nu \rho \sigma} S^{\rho \sigma}U^{\nu} \Psi^{\mu} + 2 P_{[ \mu} U_{\nu ]}\Psi^{\mu \nu}.
 \end{equation}

By taking the variation with respect to $\Psi^{\alpha}$ and $\Psi^{\alpha \beta}$ to get,

\begin{equation}
 \frac{\bar{\delta}U^{\beta}}{ \bar{\delta} S} = \frac{1}{2m} K^{\alpha}_{\mu \nu \sigma}S^{\nu \sigma} U^{\mu} ,
\end{equation}
and
\begin{equation}
 \frac{\bar{\delta}S^{\alpha \beta}}{ \bar{\delta} S} = 2 P^{[ \alpha} U^{\beta ]} .
\end{equation}

 Now, in order to obtain their corresponding deviation equation,  operate  the covariant delta derivative with respect to the path $\tau$ on both sides and using the following conditions
\begin{equation}
\frac{\bar{\delta}\Psi^{\beta}}{ \bar{\delta} S} = \frac{\bar{\delta}U^{\beta}}{ \bar{\delta} \tau}
\end{equation}
and
\begin{equation}
 \frac{\bar{\delta}}{\bar{\delta} S }\frac{\bar{\delta} {A}}{\bar{\delta} \tau }- \frac{\bar{\delta}}{\bar{\delta} \tau } \frac{\bar{\delta} {A}}{\bar{\delta} S } = K^{\alpha}_{\beta \gamma \epsilon} U^{\beta} U^{\gamma} \Psi^{\epsilon},
\end{equation}

to get

\begin{equation}
 \frac{\bar{\delta}^2 \Psi^{\beta}}{\bar{\delta} S^2} = K^{\alpha}_{\mu \nu \sigma}U^{\mu} U^{\nu}\Psi^{\sigma}  + \frac{1}{2m} (K^{\alpha}_{\mu \nu \sigma}S^{\nu \sigma} U^{\mu})_{\bar{;} \rho} \Psi^{\rho} ,
\end{equation}
and
\begin{equation}
 \frac{\bar{\delta}^2 \Psi^{\alpha \beta}}{ \bar{\delta} S^2} = S^{[ \alpha \mu} K^{\beta ]}_{\mu \nu \sigma} U^{\nu}\Psi^{\sigma} + 2(2 P^{[ \alpha} U^{\beta ]})_{\bar{;}\lambda} \Psi^{\lambda}
\end{equation}
\section{Finsler Geometry : The Cartan Approach}
It has been known that coordinates in different versions of  the Finsler space  $F(x, \dot x)$ are expressed within tangent manifolds {\bf{TM}} rather than manifolds. This may give rise to generalize the role of directional derivative $\dot x$, to become a mere  coordinate system $y^{a}$   viable for describing  gauge potentials of an  anisotropic gravitational field [5].
 Such an  issue was established by Cartan, who kept  existing  the nonlinear connection , which in return may be feasible to   express  the metric-affine gravity  theory (MAG) [15]  in its Finslerian version. The virtue of implementing  the  nonlinear connection $N^{\mu}_{\nu}$,  become responsible for for splitting the tangent bundle into two sub-bundles one of  the  horizontal coordinates and the  other is assigned for  the  vertical coordinates.  Yet, in Finsler geometry  the presence  nonlinear connection is unique, while other linear connections are many ones.  This is so valuable for describing  anisotropic  gravitational field theory . \\
The building blocks of TM are centered on considering  the adopted basis $(\delta_{\mu} , \partial_{a})$ for  the  horizontal coordinates , whose indices  appear  in  Greek letters , while the vertical ones  are expressed  in the Latin ones.\\
In Finslerian geometry there are two different types of connections linear connections as expressed in Riemannian geometry with slight modifications, and a non-linear connection $N^{a}_{k}$.  However, the latter does not be expressed in Riemannian geometry due to expressing the tangent space not on a manifold but on a tangent bundle. Thus, this space is decomposed into two a vertical space $\frac{\partial}{\partial y^{k}}$ and a horizontal space spanned by the elongated derivatives. This is can be seen as describing the nonlinear connection as gauge potential associated with its metric tensor [5] and [16]. \\
This type of vision led Vacaru to express MAG theory in Finsler space.
$$
\frac{\delta}{\delta x^{k}}= \frac{\partial}{\partial x^{k}} - N^{a}_{k}\frac{\partial}{\partial y^{k}}
$$
such that
$$
N^{a}_{k} = \frac{1}{2} g^{ab}(y^{k} \frac{\partial^{2} F^2(x^{l}, y^{l})}{\partial x^{b} \partial y^{k}}- \frac{\partial F^2(x^{l}, y^{l})}{\partial x^{b}} ).
$$
Due to the elongated derivatives the Christoffel symbol of the horizontal space is defined as
$$
\Gamma^{i}_{jk} = \frac{1}{2} g^{il}(\delta_{j} g_{kl} + \delta_{l} g_{jk} - \delta_{k} g_{lj}).
$$
 And its corresponding   vertical affine connection remains as the usual  Christoffel symbol  as given in the  Riemannian geometry.
i.e.
$$
C^{i}_{jk} = \frac{1}{2} g^{il}(\dot\partial_{j} g_{kl} + \dot\partial_{l} g_{jk} - \dot\partial_{k} g_{lj}),
$$
where $\dot\partial_{\alpha}= \frac{\partial }{\partial y^{\alpha}} .$ \\

 Accordingly, the equation of geodesic has two branches  one is for horizontal components and the other is for the vertical components as shown below

  \begin{equation}
  \frac{\nabla U^{i}}{\delta s} =0,
  \end{equation}

  where$$\frac{\nabla U^{i}}{\delta s} = \frac{\delta U^{i}}{\delta s} + \Gamma^{i}_{jk} U^{j} U^{k}  $$ as $ U^{i} = \frac{d x^{i}}{d s}$,
  and
  \begin{equation}
  \frac{D V^{a}}{D s} =0.
  \end{equation}
Provided that
$$
\frac{D V^{a}}{D s} = \frac{d V^{a}}{d s} + C^{a}_{bc} v^{b} V^{c},
$$
    and $$V^{a} =\frac{\delta y^{a}}{\delta s}.$$   This type of choice may be appropriate to express bi-gravity geometrically , which  will be studied in our future work.
\subsection{Metric Structure and Non-linear Connection}
It is well known that, the metric structure  for a combined space $V^{8}$ of both vertical and horizontal coordinates is defined using the non-linear connection  [5]. Such a connection may act as gauge potentials to express the case of a metric of anisotropy  features in the following way.
\begin{equation}
 g_{M N} =\left(
  \begin{array}{cc}
    g_{\mu \nu} + h_{a b} N^{a}_{\mu}N^{b}_{\nu}  & N_{b \mu} \\
    N_{a \mu} & h_{ab} \\
  \end{array}
\right)
\end{equation}
such that $\mu , \nu =0,1,2,3$ and $ m, n = 4,5,6,7$ where
$N^{\alpha}_{\beta}$  acting as gauge potentials.\\
To define 8-dimensional metric
\begin{equation}
dS^2 = G_{MN} d X^{M} \bigotimes dX^{N}.
\end{equation}
i.e.
\begin{equation}
  ds^2 = g_{ij}(x,y) dx^{i}dx^{j} + {h_{ab}}(x,y) \delta y^{a} \delta^{b},
\end{equation}
  where $\delta y^{a}$ is the elongated derivative as defined in the adopted basis $(\partial, d)$.   In this case, the manifold is split into vertical and horizontal coordinates subject to an  adopted basis $(\delta, \partial)$ due to the existence of nonlinear connections.

Moreover, the spray $G^{\mu}$ is connected with a nonlinear connection $N^{\alpha}_{\beta}$ in which
\begin{equation}
 G^{\mu} = \frac{1}{4} N^{\alpha}_{\beta} \dot{x}^{\beta}
\end{equation}
i.e. replacing $ y ^{\mu} = \dot{x}^{\mu}$ .
\begin{equation}
N^{\alpha}_{\beta}  = 2 \frac{\partial G^{\alpha}}{\partial \dot{x}^{\beta}}.
\end{equation}
The importance of the non-linear connection is causing the ability to decompose the tangent space to the tangent bundle at point $ (x, \dot{x})$ into a vertical span $ \frac{\partial}{\partial \dot{x}}$ and a horizontal elongated derivative
\begin{equation}
\frac{\delta}{\delta x^{\mu}} = \frac{\partial}{\partial x^{\mu}} - N^{\nu}_{\mu} \frac{\partial}{\partial \dot{x}^{\nu}}.
\end{equation}
A non linear curvature is defined as
\begin{equation}
 R^{\mu}_{\nu \sigma} = \frac{\delta N^{\mu}_{\sigma}}{\delta \dot{x}^{\nu}} - \frac{\delta N^{\mu}_{\nu}}{\delta \dot{x}^{\sigma}}.
\end{equation}
 This may lead us consider some of them as an illustration to this claim.

Such an explanation may be an advantage to  describe the micro-internal  behavior of particles in Finsler space, [16]. The  y-vector can also be regarded to measure the fluctuations of space-time. This can be seen in terms of $C_{ijk}$ or its associated curvature $S^{i}_{jkl}$. \\

 Accordingly, the coordinate system of the tangent bundle has to be split into two categories horizontal and vertical ones. The causality of this performance is related to identifying Finsler geometry keeping the same behavior as of Riemannian geometry to express bigravity theory using Finslerian geometry [17].

\subsection{ Geodesic and Geodesic Deviation in h-derivative and v-derivative}
Paths of test particles in Finsler geometry, may be determined by applying the action principle on  the following Lagrangian function in the following manner,
\begin{equation}
L= g_{\mu \nu} (x, y) U^{\mu} U^{\nu} + h_{a b}(x,y) V^{a} V^{b}.
 \end{equation}

Taking the variation with respect to $U^{\alpha}$ and $V^{c}$ respectively, we obtain the following set of geodesics: (38) and (39)\\

Thus, if we apply (23)  on both (33) and (34), we obtain their corresponding deviation equations.\\
For h-dervative: 
\begin{equation}
\frac{\nabla^2 \Psi^{\alpha}}{\nabla s^2} = \bar{R}^{\alpha}_{\beta \gamma \delta} U^{\beta} U^{\gamma} \Psi^{\delta}
\end {equation}

where $ \bar{R}^{\alpha}_{\beta \gamma \delta} = \bar{\Gamma}^{\alpha}_{\beta \delta || \gamma } - \bar{\Gamma}^{\alpha}_{\beta \gamma || \delta } + \bar{\Gamma}^{\lambda}_{\beta \delta }\bar{\Gamma}^{\alpha}_{\lambda \gamma } - \bar{\Gamma}^{\lambda}_{\beta \gamma }\bar{\Gamma}^{\alpha}_{\lambda \delta }.   $ \\
And for v-derivative: 
\begin{equation}
\frac{D^2 \Phi^{\alpha}}{D s^2} =  C^{\alpha}_{\beta \gamma \delta} V^{\beta} V^{\gamma} \Phi^{\delta}
\end{equation}
where,
$${C}^{\alpha}_{\beta \gamma \delta} = {\Gamma}^{\alpha}_{\beta \delta | \gamma } - {\Gamma}^{\alpha}_{\beta \gamma | \delta } + {\Gamma}^{\lambda}_{\beta \delta }{\Gamma}^{\alpha}_{\lambda \gamma } - {\Gamma}^{\lambda}_{\beta \gamma }{\Gamma}^{\alpha}_{\lambda \delta }.   $$

\section{Spinning equation of Finsler-Cartan Approach}
In a similar way as presented in section 3 , we are going to derive the spinning equations for short, using the following transformation
Using the previous technique to derive the spin equations ,for short, from geodesic and geodesic deviation equations, we suggest the following functions:
\begin{equation}
 \bar{U} = U + \beta\frac{\nabla \Psi^{\alpha}}{\nabla s},
\end{equation}
and
\begin{equation}
 \bar{V} = V + \beta\frac{D\Phi^{\alpha}}{D s}.
\end{equation}

Differentiating both sides of (45)and (46) with respect to $\frac{\nabla}{\nabla s} $ and $\frac{D}{D s}$ respectively, and using (43) and (44) we obtain,

\begin{equation}
 \frac{\nabla {\bar U}^{\beta}}{ \nabla s} = \frac{1}{2m} R^{\alpha}_{\mu \nu \sigma}S^{\nu \sigma} U^{\mu} ,
\end{equation}
and
\begin{equation}
 \frac{D {\bar V}^{\beta}}{ D s} = \frac{1}{2m} S^{\alpha}_{\mu \nu \sigma}\bar{S}^{\nu \sigma} V^{\mu} .
\end{equation}
\subsection{Spinning and Spinning Deviation equation Using Finsler-Cartan Approach}
{\underline{{i} Case $ P^{\mu} = m U^{\mu}$ and $ \tilde{P}^{a} = m V^{a}$}}
\begin{equation}
L= g_{\mu \nu} (x) U^{\mu} \frac{\nabla \Phi^{\nu}}{\nabla s} + S_{\mu \nu}\frac{\nabla \Psi^{\mu \nu}}{\nabla s} + h_{a b}(y) V^{a} \frac{D \Phi^{b}}{Ds}+ \tilde{S}_{ab} \frac{D \Phi^{a b}}{D s} + \frac{1}{2m} R_{\mu \nu \rho \sigma} S^{\rho \sigma}U^{\nu} \Psi^{\mu} + \frac{1}{2m} S_{a b c d} \tilde{S}^{c d}U^{b} \Psi^{a}.
 \end{equation}
Now,  taking the variation with respect to $\Psi^{\alpha}$ and $\Psi^{\alpha \beta}$ to obtain

\begin{equation}
 \frac{\nabla {\bar U}^{\beta}}{ \nabla s} = \frac{1}{2m} R^{\alpha}_{\mu \nu \sigma}S^{\nu \sigma} U^{\mu} ,
\end{equation}
 Thus, In a similar way we obtain their corresponding deviation equations  .

 For h-components:
\begin{equation}
 \frac{\nabla^2 \Psi^{\beta}}{\nabla s^2} = R^{\alpha}_{\mu \nu \sigma}U^{\mu} U^{\nu}\Psi^{\sigma}  + \frac{1}{2m} (R^{\alpha}_{\mu \nu \sigma}S^{\nu \sigma} U^{\mu})_{|| \rho} \Psi^{\rho} ,
\end{equation}
and

\begin{equation}
 \frac{\nabla^2 \Psi^{\alpha \beta}}{ \nabla s^2} = S^{[ \alpha \mu} R^{\beta ]}_{\mu \nu \sigma} U^{\nu}\Psi^{\sigma}   ,
\end{equation}

 While,for v-components:\\
By taking the variation with respect to $\Phi^{a}$ and $\Phi^{a b}$

\begin{equation}
 \frac{D \bar{V}^{a}}{ D s} = \frac{1}{2m} S^{a}_{ b c d }\bar{S}^{c d } V^{b} .
\end{equation}
and
\begin{equation}
 \frac{D \tilde{S}^{a b}}{ D s^2} = 0  .
\end{equation}

Using the similar technique for obtaining spinning deviation equations we get,

\begin{equation}
 \frac{D {\Phi}^{a b }}{ D s} = \frac{1}{2m} S^{\alpha}_{\mu \nu \sigma}\bar{S}^{\nu \sigma} V^{\mu} ,
\end{equation}

\begin{equation}
 \frac{D^2 {\bar{\Phi}^{a}}}{D s^2} = S^{a}_{b c d }V^{b} V^{c}\Phi^{d} + \frac{1}{2m} ( {S}^{a}_{b c d} \tilde{S}^{c d } V^{b} )_{; e} \Phi^{e} ,
\end{equation}

and
\begin{equation}
 \frac{D^2 \Phi^{a b}}{ D s^2} = S^{[ a c} S^{b ]}_{c d e} U^{d}\Psi^{e}   .
\end{equation}

{\underline{{ii} Case $ P^{\mu} \neq m U^{\mu}$ and $ \tilde{P}^{a} \neq m V^{a}$}} \\
In this case we obtain the spinning  and spinning deviation equations  for objects  having some intrinsic properties due to  regarding the momentum is not solely the product of mass times for vector velocity  [14]
\begin{equation}
L= g_{\mu \nu} (x) P^{\mu} \frac{\nabla \Phi^{\nu}}{\nabla s} + S_{\mu \nu}\frac{\nabla \Psi^{\mu \nu}}{\nabla s} + h_{a b}(y) \tilde{P}^{a} \frac{D \Phi^{b}}{Ds}+ \tilde{S}_{ab} \frac{D \Phi^{a b}}{D s} + \frac{1}{2} R_{\mu \nu \rho \sigma} S^{\rho \sigma}U^{\nu} \Psi^{\mu} + \frac{1}{2} S_{a b c d} \tilde{S}^{c d}U^{b} \Psi^{a} + 2P_{[ \mu} U_{\nu ]}\Psi^{\mu \nu} + 2\tilde{P}_{[ \mu} V_{\nu ]}\Phi^{\mu \nu}
 \end{equation}
Thus,  for h-components,  we take  the variation with respect to $\Psi^{\alpha}$ and $\Psi {\alpha \beta}$

\begin{equation}
 \frac{\nabla { P}^{\alpha}}{ \nabla s} = \frac{1}{2} R^{\alpha}_{\mu \nu \sigma}S^{\nu \sigma} U^{\mu} ,
\end{equation}
and
\begin{equation}
 \frac{\nabla {S}^{\alpha \beta}}{ \nabla s} = 2 P^{[ \alpha} U^{\beta ]} ,
\end{equation}

In a similar way we obtain their corresponding deviation equations  .

\begin{equation}
 \frac{\nabla^2 \Psi^{\beta}}{\nabla s^2} = R^{\alpha}_{\mu \nu \sigma}P^{\mu} U^{\nu}\Psi^{\sigma}  + \frac{1}{2} (R^{\alpha}_{\mu \nu \sigma}S^{\nu \sigma} U^{\mu})_{|| \rho} \Psi^{\rho} ,
\end{equation}
and

\begin{equation}
 \frac{\nabla^2 \Psi^{\alpha \beta}}{ \nabla s^2} = S^{[ \alpha \mu} R^{\beta ]}_{\mu \nu \sigma} U^{\nu}\Psi^{\sigma} +  2  (P^{[ \alpha} U^{\beta ]})_{| \rho} \Psi^{\rho}  .
\end{equation}

And for v-coordinates, we can obtain
\begin{equation}
 \frac{D \bar{P}^{a}}{ D s} = \frac{1}{2} S^{a}_{ b c d }\bar{S}^{c d } V^{b} .
\end{equation}
and
\begin{equation}
 \frac{D \tilde{S}^{a b}}{ D s^2} = 2 P^{[a}U^{b ]}   .
\end{equation}

Using the similar technique for obtaining spinning deviation equations  to get

\begin{equation}
 \frac{D {\Phi}^{a b }}{ D s} = \frac{1}{2m} S^{\alpha}_{\mu \nu \sigma}\bar{S}^{\nu \sigma} V^{\mu} .
\end{equation}

\begin{equation}
 \frac{D^2 {\bar{\Phi}^{a}}}{D s^2} = S^{a}_{b c d }\tilde{P}^{b} V^{c}\Phi^{d} + \frac{1}{2} ( {S}^{a}_{b c d} \tilde{S}^{c d } V^{b} )_{; e} \Phi^{e} .
\end{equation}

and
\begin{equation}
 \frac{D^2 \Phi^{a b}}{ D s^2} =  S^{[ a c} S^{b ]}_{c d e} V^{d}\Phi^{e} + 2 ( P^{[a}V^{b ]})_{; d} \Phi^{d}.
\end{equation}

Due to the role of the adopted basis $(\delta, \partial)$, the two equations of spinning motion look independently.  From this perspective,  there is   no interacting terms between  $U^{\alpha}$ and $V^{\alpha}$ , by  considering  $y^{\alpha} = \frac{dx^{\alpha}}{ds}$ .  This means that we obtain the associated rate of change of  spinning acceleration  i.e. the spinning jerk.
This may have an impact on adjusting trajectories of space navigation .
\section{Applications of Spinning Equations in Finsler Spaces}
\subsection{Finslerian Space: The Brandt Approach}
Brandt [18] has suggested the following square  line element to obtain the geodesic in Finslerian fields, provided that he  discarded the effect of non-linear connection, to express the Lagrangian function as combined between velocity and acceleration, in our present work we are going to extend his vision to to obtain the associate equations of spinning and rate of change of spinning (spinning jerk) related to proper maximal acceleration  .
\begin{equation}
d \sigma^{2} = g_{\mu \nu} U^{\mu} U^{\nu}  + \rho_{0} DU^{\mu} DU^{\nu}
\end{equation}
where $ D U^{\mu} = d U^{\mu} + \cs{\lambda}{\nu}{\mu} U^{\lambda} U^{\nu}$ .\\etry
Thus it corresponding Lagrangian Function becomes
\begin{equation}
 L_{\sigma} = g_{\mu \nu} U^{\mu} U^{\nu}  + \rho_{0}g_{\mu \nu} \frac{D U^{\mu}}{ D \sigma} \frac{D U^{\nu}}{ D \sigma}
\end{equation}
where $\rho_{0}$ is the proper maximal acceleration [19]. \\
Taking the variation with respect to $x^{\mu}$ and ${v}^{\alpha}$ one obtains

\begin{equation}
\frac{D U^{\alpha} }{D \sigma} =0
\end{equation}
and

\begin{equation}
\frac{D^2 U^{\alpha} }{D \sigma^2} =0.
\end{equation}
Thus, if we suggest its Bazanski Lagrangian to become

\begin{equation}
 L_{\sigma} = g_{\mu \nu} U^{\mu} \frac {D \Psi^{\mu}}{D \sigma}  + \rho_{0} g_{\mu \nu} \frac{D U^{\mu}}{ D \sigma} \frac{D \Phi^{\mu}}{ D \sigma}
\end{equation}
where $ \Phi^{\mu}$ is the corresponding deviation vector.

Accordingly,  taking the variation with respect to $\Psi^{\alpha}$ and $\Phi^{\alpha}$ as$  \Phi^{\beta} = \frac{D \Psi^{\beta}}{D s}$ simultaneously, we obtain  equations  expressed in  ( 33) and (34) respectively. \\
 Following the same technique as mentioned in sections (…). Yet, we  obtain their corresponding deviation equations
 \begin{equation}
 \frac{D^{2} \Psi^{\mu}}{ D \sigma^2} = R^{\mu}_{\nu \rho \delta} U^{\nu} U^{\rho} \Psi^{\delta}
\end{equation}
and
\begin{equation}
 \frac{D^{2} \Phi^{\mu}}{ D \sigma^2} = R^{\mu}_{\nu \rho \delta} \frac{D{v}^{\nu}}{D \sigma} \frac{D v^{\rho}}{D \sigma} \Phi^{\delta}.
\end{equation}
Also, in a similar way as the previous sections,  we  derive their corresponding spinning equations to become in the following way:\\
we assume the following Lagrangian
$$
L= g_{\mu \nu} P^{\mu} \frac{D \Psi^{\nu}}{D \sigma} + S_{\mu \nu}\frac{D \Psi^{\mu \nu}}{D \sigma} + \frac{1}{2}R_{\mu \nu \rho \sigma}U^{\nu}S^{\rho \sigma}\Psi^{\mu} + 2 P_{[\mu}  U_{\nu ]} \Psi^{\mu \nu}
$$
 \begin{equation}
 +  \rho_{0} g_{\mu \nu} \frac{D P^{\mu}}{D \sigma } \frac{D \Phi^{\nu}}{D \sigma}
+ \frac{1}{2}R_{\mu \nu \rho \sigma}\frac{d U^{\nu}}{d \sigma}\frac{D S^{\rho \sigma}}{D \sigma}\Psi^{\mu} + 2 \frac{d P_{[\mu}}{d \sigma}  \frac{ d U_{\nu ]}}{d \sigma} \Psi^{\mu \nu}.
\end{equation}
Also, taking the variation with respect to $\Psi^{\alpha}$ and $\Psi^{\alpha \beta}$ to get: \\
For h-coordinates
\begin{equation}
 \frac{D P^{\mu}}{ D \sigma} = \frac{1}{2} R^{\mu}_{\nu \rho \delta} v^{\nu} S^{\rho \delta}
\end{equation}
and
\begin{equation}
 \frac{D S^{\mu \nu}}{ D \sigma} =  2 P^{[ \mu} v^{\nu ]}.
\end{equation}
While, for v-coordinates \\
we take the variation with respect to $\Phi^{\alpha}$ and $\Phi^{\alpha \beta}$ to obtain
\begin{equation}
\frac{D^{2} P^{\mu} }{D \sigma^{2} } = \frac{1}{2} R^{\mu}_{\nu \rho \delta} \frac{D U^{\nu} }{D \sigma} \frac{D S^{\rho \delta}}{D \sigma}
\end{equation}
and
\begin{equation}
 \frac{D^2 S^{\mu \nu}}{ D \sigma^2} =  2 \frac{ D^{[ \mu}{P}}{D \sigma} \frac{D U^{\nu ]}}{D \sigma}.
\end{equation}

Nevertheless, the above tangent bundle space can be expressed in Kaluza-Klein space to be considered as follows of 8-dimensions whose line element is expressed as follows [20]
\begin{equation}
d \epsilon^{2}  =  G_{M N}(x, v)  d x^{M} dx^{N} , {M, N  = 0,1,2..7}
\end{equation}
where the bundle coordinates are ${x^{M}} \equiv x^{\mu}, \rho_{0} v^{\mu}$ such that the metric of the tangent bundle of space-time is
\begin{equation}
 G_{M N} =\left(
  \begin{array}{cc}
    g_{\mu \nu} + g_{\alpha \beta} A^{\alpha}_{\mu}A^{\beta}_{\nu}  & A_{n \mu} \\
    A_{m \mu} & g_{mn} \\
  \end{array}
\right)
\end{equation}
such that $\mu , \nu =0,1,2,3$ and $ m, n = 4,5,6,7$ where
$$
A^{\mu}_{\nu} = \rho v^{\lambda} \cs{\lambda}{\nu}{\mu}
$$
which act as gauge potentials [18].\\
These above equations of geodesic  may  be obtained using another  Lagrangian
\begin{equation}
L_{KKB} =  G_{MN} U^{M} \frac{D \Psi{N}}{D \sigma}
\end{equation}
to give
\begin{equation}
\frac{D U^{M}}{D \epsilon} =0
\end{equation}
and
\begin{equation}
\frac{D^2 Psi^{M}}{D \epsilon^{2} } = R^{M}_{NAB} U^{N} U^{A} \Psi^{B}.
\end{equation}

While, their corresponding spinning equation become the same as the Papapetrou equations in 8-dimensions [20]  i.e.
\begin{equation}
\frac{D P^{M}}{D \epsilon} = R^{M}_{NOP} \frac{dx^{N}}{d \epsilon}S^{OP}
\end{equation}
and
\begin{equation}
\frac{D S^{MN}}{D \epsilon} = 2 P^{[M} \frac{d x^{N]}}{ d \epsilon} .
\end{equation}
Thus,  the two separate equations are expressed in Kaluza-Klein space as one geodesic equation the first four components are expressing acceleration vector,while the last ones are defining the jerk vector.
\subsection{On the relation between Spin Tensor and Proper Maximal Acceleration}
Brandt [20] showed a relationship between the proper maximal acceleration $\rho_{0}$ and the Riemann-Chrstofel Tensor in the following way,
\begin{equation}
\rho^{2}_{0} = R_{\alpha \beta \gamma \sigma} U^{\beta} U^{\gamma} N^{\alpha} N^{\sigma},
\end{equation}
In which ,  the  corresponding  unit vector  is defined as $ N^{\alpha} N_{\alpha} =-1$ . \\
If we assume that N -vector with deviation vector in the following way
\begin{equation}
N^{\alpha} = \frac{s}{m} \Psi^{\alpha},
\end{equation}
then,  we find out that,
\begin{equation}
\rho^{2}_{0} =  (\frac{s}{m})^2 R_{\alpha \beta \gamma \sigma} U^{\beta} U^{\gamma} \Psi^{\alpha} \Psi^{\sigma}.
\end{equation}
Consequently,
\begin{equation}
\rho^{2}_{0} =  \frac{1}{2m} R_{\alpha \beta \gamma \sigma} U^{\beta} S^{\gamma \sigma} N^{\alpha}.
\end{equation}
 Meanwhile, using the Papapetrou equation [14], we find that
\begin{equation}
\rho^{2}_{0} =  \frac{DU_{\alpha}}{Ds} N^{\alpha}.
\end{equation}
Yet, multiplying both sides by $N_{\epsilon}$, provided that $N^{\alpha}N_{\epsilon} = \delta^{\alpha}_{\epsilon}$ we obtain the relationship between the covariant acceleration and proper maximal acceleration [20]
\begin{equation}
\frac{DU^{\alpha}}{Ds} = \rho^2_{0} N^{\alpha}.
\end{equation}
Using equations (29) and  (89)in (90 ) we get
\begin{equation}
\frac{D\Psi^{\alpha}}{D\tau} = \frac{s}{m} \rho^2_{0} \Psi^{\alpha}.
\end{equation}
 If we Operate the covariant derivatives on both sides of (93) to get
  \begin{equation}
\frac{D^{2} \Psi^{\alpha}}{D\tau^2} =  (\frac{s \rho_{0}}{m})^{2} \rho^2_{0} \Psi^{\alpha}.
\end{equation}
The latter gives a significance of a wave equation regulated by the quantity of $( \frac{s \rho_{0}}{m})^{2}$ , from which we can relate between the deviation vector and the proper maximal acceleration with the ration between the magnitude of spinning object and its mass.

\section{Conclusion }
  In this study, we have obtained the set of spinning and spinning deviation equations for the Riemannian analog in Finslerian geometry  as clarified in  the Rund and Cartan approaches. In this study,  we have  obtained sets of spinning and spinning deviation equations to examine the stability of any rotating object on its orbit. This study is in favor of the argument of Argzishiev and Dzhunushaliev [21] who  emphasized  the gyroscopic precession in Finsler is different from the usual Riemannian treatment. Meanwhile, it is quite evident to note that the behavior of the  spin connection as gauge potential for a covariant derivative under general coordinate transformation and local Lorentz transformation in the  Riemannian  geometry [22] may be equivalent to the non- linear connection in the Finsler geometry .
   Owing to this analogy,  we reach to regard the behavior of the spin connection in non- Riemannian geometries as the untested effect of nonlinear connection in these types of geometries. Such a result will be extended to examine its role in torsion and curvature, having a Finslaerian flavor [23] on the spinning motion and the effect of their corresponding spinning jerk in our future  work.

\section*{References}
{[1]} S. F. Rutz, and F. M. Pavia,  Finslerian Geometry, Ed. P.L. Antonelli, Kluwer Academic Publisher (2000), \\
{[2]} C.C. Perelman, PAIJ, vol. 2, issue 6 (2018), \\
{[3]} H. Rund,{\it{The Differential Geometry of Finsler Spaces}}, Springer-Verlag (1959),\\
{[4]} M. Anastasici and H. Shimada,  Finslerian Geometry, Ed. P.L.  Antonelli, Kluwer Academic Publisher (2000),  \\
{[5]}  S.I. Vacaru, hep-th/0310132 (2003), \\
{[6]} S. L.  Bazanski  {J. Math. Phys.} {\bf {30}}, 1018 (1989), \\
{[7]} M.E. Kahil,  J. Math. Phys. {\bf {47}},052501 (2006), \\
{[8]} Magd E. Kahil , Gravit. Cosmol.,  vol.{\bf{24}}, issue 1, 83 (2018),  \\
{[9]} Magd E. Kahil , ADAP {\bf{3}}, 136 (2018) \\
{[10]} Magd E. Kahil, Gravit. Cosmol. {\bf{25}}, vol.3,  268 (2019) \\
{[11]}M. Pavsic and Magd E. Kahil Open Physics {\bf{10}}, 414 (2012). \\
([12])M. Heydrai-Fard, M. Mohseni, and  H.R. Sepanigi {Phys. Lett. B, {\bf{626}}, 230 (2005), \\
{[13]} D. Bini and A Geralico, Phys. Rev. D, {\bf{{84}}},104012; arXiv: 1408.4952 (2011),\\
{[14]} A. Papapetrou , Proceedings of Royal Society London A, {\bf{209}} , 248(1951), \\
{[15]}  F. W. Hehl, J.D. Mc Grea, E.W. Mielke and Y. Ne'emann Phys Rep. {\bf{258}}, 1 (1995), \\
{[16]} R. Miron and M. Anastasici, {\it{ Vector Bundles and Lagrange Spaces}}, Blakan Press (1997), \\
{[17]} J. Sklakla, and M. Visser,  arXiv: 1008.0689 (2010). \\
{[18]} H. Brandt , Foundations of Physics Letters, vol.4, Number 6 (1991), \\
{[19]} C. Castro.  hep-th/ 0211053 (2002)  \\
{[20]} H. Brandt, Finslerian Geometry, Ed. P.L. Antonelli, Kluwer Academic Publisher (2000),  \\
{[21]} A. K. Argauzin, V. Dzhunsaliev, gr-qc/1303.0912 (2013), \\
{[22]} P.D. Collins, A.D. Martin,  and E.J. Squires {\it{ Particle Physics And Cosmology}} John Wily and Sons (1989), \\
{[23]} M. I. Wanas, M. E. Kahil and Mona M. Kamal , Gravit Cosmol. vol.{\bf{22}}, issue 4, 345  (2016),\\

\end{document}